\documentclass{article}

\usepackage{arxiv}

\usepackage[utf8]{inputenc} % allow utf-8 input
\usepackage[T1]{fontenc}    % use 8-bit T1 fonts
\usepackage{hyperref}       % hyperlinks
\usepackage{url}            % simple URL typesetting
\usepackage{booktabs}       % professional-quality tables
\usepackage{amsfonts}       % blackboard math symbols
\usepackage{nicefrac}       % compact symbols for 1/2, etc.
\usepackage{microtype}      % microtypography
\usepackage{lipsum}
\usepackage{graphicx}
\graphicspath{ {./images/} }

\title{Adding web pentesting functionality to PTHelper}

\author{
María~Olivares~Naya \\
Universidad Carlos III de Madrid\\
Avda de la Universidad 30\\
Legan\'es, Madrid, Spain\\
\texttt{100428974@alumnos.uc3m.es},
\And
Jacobo Casado de Gracia \\
jtsec \\
Avenida de la Constitución nº 20 Oficina 208 \\
Granada, Spain\\,
\And 
Alfonso S\'anchez-Maci\'an \\
Universidad Carlos III de Madrid\\
Avda de la Universidad 30\\
Legan\'es, Madrid, Spain\\
\texttt{alfonsan@it.uc3m.es},
}

\begin{document}
\maketitle
\begin{abstract}
Web application pentesting is a crucial component in the offensive cybersecurity area, whose aim is to safeguard web applications and web services as the majority of the web applications are mounted in publicly accessible web environments. This method requires that the cybersecurity experts pretend and act as real attackers to identify all the errors and vulnerabilities in web applications with the objective of preventing and reducing damages. As this process may be quite complex and the amount of information pentesters need may be big, being able to automate it will help them to easily discover the vulnerabilities given.  This project is the direct continuation of the previous initiative called "PThelper: An open source tool to support the Penetration Testing process". This continuation is focused on expanding PThelper with the functionality to detect and later report web vulnerabilities in order to address emerging threats and strengthen organizations’ ability to protect their web applications against potential cyber-attacks.

\end{abstract}


\section{Introduction}
\label{sec:introduction}

In today’s digital age,  the use of the internet and online services are becoming more and more integral, and web applications play a key role in our daily lives. However, this presence in our lives can lead to significant cybersecurity risks. Web applications can have different vulnerabilities that can be exploited by cybercriminals, giving them a free hand to access sensitive information, compromise the integrity of data, or even take control of the whole system.  Due to this situation, the concept of Penetration Testing was born as an essential component in the security and protection of web applications and their users.

Ministry of Interior announced that in Spain during 2022 (through the last official cybercrime report \cite{cybercrime}) the number of cyberattacks increased 22,7\% over the previous year, reaching 374.737 cybercrimes. In 2023, the number of cyberattacks did not decrease either and were produced by misconfigured, outdated systems or using stolen, leaked or randomly tested credentials with brute force attacks\cite{cabrera-2024}.

These cyberattacks were not targeted to a specific group or sector, public bodies and entities such as the Tax Agency or the Ministry of Interior, and private entities such as Telepizza or Air Europa\cite{aireuropa} have been the target of cyberattacks, resulting in large losses. 

Due to the increase in data breaches and attacks, organizations are implementing security measures to safeguard systems to protect their assets and personnel, including offensive security actions such as penetration testing, also known as \textit{pentesting}.

The penetration testers are computer security specialists whose goal is to identify flaws in an organization's infrastructure and web applications by emulating the tactics and techniques of a potential attacker. These tactics or tests, known as (web)pentesting, aim to discover and remediate flaws before a real attacker can discover and exploit them, thus minimizing the risk of security breaches.

In this paper, it is explained the new functionalities of PThelper\cite{degracia2024pthelper} for discovering web vulnerabilities. These functionalities have been designed and parametrized in a modular way, allowing an easy adaptation and the possibility of changing the tools used, guaranteeing the versatility and longevity of PThelper. Moreover, PThelper has been parallelizedto improve efficiency.

This paper is divided into 5 different parts: 
\begin{itemize}
    \item Background on web \textit{pentesting}: Provides a context over the techniques and methodologies used in web \textit{pentesting}
    \item Architecture of the project:  Explains its architecture and implementation
    \item Tool usage and information: Explain the language and packages used, how the program must be executed, and the commands and arguments used for it.
    \item Testing: This section deals with the tests carried out to check the effectiveness of the system. 
    \item Conclusions:  Present the conclusions derived from the development of this work, highlighting the achievements made, the limitations encountered, and possible future directions for the improvement and expansion of the system.
\end{itemize}
\section{Background on web \textit{pentesting}}
\subsection{Web Vulnerabilities}
A web vulnerability is a weakness or a flaw in a system, application, or web service that can be exploited by an attacker in order to compromise the security of the system or to gain unauthorized access to sensitive information. These vulnerabilities arise due to design, implementation or configuration errors, and if the attackers discover any, they can perform a variety of malicious actions such as performing denial-of-service (DoS) attacks, executing arbitrary code, stealing sensitive information, modifying or deleting data, among others. Therefore, if applications have web vulnerabilities and are discovered, they can have serious consequences for the security and integrity of the whole system that is in charge of executing the web application. It is really important for developers and system administrators to proactively identify and remediate these vulnerabilities to protect their digital assets against potential attacks.

Two of the most common and relevant web vulnerabilities are SQL injection (SQLi) and Cross-Site Scripting (XSS). Both vulnerabilities present a significant risk for web application security and can have devastating consequences if they are not properly managed \cite{ccnstic812}.
\subsubsection{SQL Injection}
SQL injection is a web vulnerability that affects to web applications that interact with databases using SQL. This vulnerability was discovered in the 1990s when web applications started using databases to store and retrieve information dynamically. This web vulnerability was a considerable problem due to the lack of proper validation of user input in the SQL queries, letting the attacker to manipulate the queries and obtain unauthorized access to the database or perform actions without the right permissions.

Later time, in the 2000s, the concept of SQL injection was popularized when the successful attacks (that compromised important web systems) came to the fore. Since that moment, SQL injection has been one of the most common and dangerous vulnerabilities, and responsible for a great number of attacks and data breaches.

To prevent SQL injection developers must implement  secure coding practices like using parametrized queries or properly validating the user input.

One example of SQL injection attack is the one of Yahoo in 2012, D33Ds Company hackers carried out a SQL injection attack exploiting a vulnerability in one of Yahoo's subdomains, Yahoo service, but Yahoo did not publish the parameters of the vulnerable subdomain to avoid greater evils as with that attack 450.000 accounts were compromised and published on the Internet
\subsubsection{Cross-Site Scripting}
As SQL injection, Cross-Site Scripting is a web vulnerability that was discovered in the 1990s when web applications started using dynamic markup languages, such as HTML and JavaScript, to create interactive and dynamic content on Web pages.

Cross-Site Scripting is a web vulnerability that lets the attacker inject and execute malicious JavaScript code in web applications and doing that, being able to steal session cookies, redirect a user to malicious web applications or execute nonauthorized actions. Similar to SQL injection, Cross-site Scripting has been one of the most common and dangerous vulnerabilities, and is responsible for a great number of attacks and data breaches.

To prevent Cross-site Scripting developers must implement secure coding practices like properly validating user input and implementing specific mitigations for this vulnerability, such as the Content Security Policy, which is a security mechanism to protect websites from these types of attacks by specifying directives that the browser must follow.

\subsection{OWASP TOP 10}
OWASP Top 10, created by Open Web Security Project (OWASP), is a list of the 10 most critical security vulnerabilities in web applications. With this list, pentesters can see the most common threats and how to protect against them, thus improving the overall security of web applications.

According to the list updated in 2021, both vulnerabilities described above are in the third position (injection). Furthermore, in the 2017 list both vulnerabilities also appeared, SQL injection in first place (injection) and XSS in seventh place (XSS, as they were previously in different categories)\cite{10}.

\subsection{testssl and sslscan}
\textit{testssl}\cite{testssl} and \textit{sslscan}\cite{sslscan} are two popular security evaluation tools used to perform security testing on SSL/TLS servers.

In terms of functionality, sslscan is more focused on scanning and enumerating the characteristics of the SSL/TLS protocol of a server, performing basic tests, and providing information on ciphersuites, protocols, and some known vulnerabilities while testssl is a more complete option as it provides a wider range of tests, information and security evaluations.

In terms of output, sslscan has a more concise output, providing a general vision of the SSL/TLS server's characteristics meanwhile testssl has a more detailed and descriptive output including information of each test and details of the configurations, protocols, certificates and vulnerabilities.

Lastly, in terms of usability sslscan is faster and easier to use so it is ideal for fast scans and retrieving basic information meanwhile testssl is more complex and slower, but retrieves more information about the TLS implementation.

Despite the features and advantages explained above, both have a significant limitation: the lack of interpretation of the results. Both testssl and sslscan generate detailed reports on the SSL/TLS configuration but they only provide technical data without offering a clear interpretation of the results and for a complete and accurate security assessment it is essential to generate the technical reports but also to interpret them in the context of up-to-date norms and standards.

\subsection{SQLMap and SQLMapAPI}
SQLMap\cite{sqlmap} is an open source tool used for detecting and exploiting SQL injection vulnerabilities in web applications automatically, letting the pentesters test the security of your databases and correct problems before they can be exploited by malicious attackers.\newline 

Features of SQLMap:
\begin{itemize}
    \item Vulnerabilities detection: SQLMap can detect a great amount of SQL injection types including error based, time based and blind injections
    \item Automatic exploitation: SQLMap lets the automatic exploitation of the discovered vulnerabilities to extract SENSIBLE information such as names of tables, columns and data.
    \item Multiple Database Support: SQLMap supports a variety of databases such as MySQL, PostgreSQL, Oracle, Microsoft SQL Server and SQLite.
    \item Injection techniques: SQLMap uses several injection techniques to ensure that all potential vulnerabilities are detected
    \item Proxy support: Proxies can be used for testing in order to avoid detection and allow testing from different locations.
\end{itemize}
SQLMapAPI is a RESTful API for SQLMap, which allows users to interact with SQLMap via HTTP requests. SQLMapAPI is used to integrate SQLMap into web applications or for use in environments where automation and integration with other systems are essential. \newline

Features of SQLMapAPI:
\begin{itemize}
    \item RESTful Interface: SQLMapAPI provides an HTTP-based API interface to interact with SQLMap.
    \item Automation and Integration: Facilitates the integration of SQLMap into CI/CD systems, web applications, and other security tools via API calls.
    \item Remote Control: Enables remote control of SQLMap instances, which is useful for managing security testing from a centralized location.
\end{itemize}

\subsection{XSpear}
XSpear\cite{xspear} is an automatic and advanced tool that is used to detect and exploit cross-site scripting (XSS) vulnerabilities in web applications. This open-source tool written in Ruby provides an intuitive and powerful interface for identifying and analyzing different XSS attacks. \newline

Features of XSpear:
\begin{itemize}
    \item  XSS Test Automation: XSpear is able to detect XSS vulnerabilities in an automatic way, reducing the time and effort required for manual tests.
    \item Attack modes: XSpear offers different attack modes, including Reflected, Stored and DOM XSS attacks.
    \item Generation of Custom Payloads: XSpear is able to generate and use custom payloads for testing the web application to different types of XSS attacks.
    \item Response: XSpear provides a response report detailing the vulnerabilities found with a description of the vulnerability, the  and payload used along with the \textbf{possible mitigations.}
    \item Integration with CI/CD: XSpear can be integrated with continuous integration and continuous delivery (CI/CD) systems, enabling the automation of security testing.
    \item Easy to use: XSpear is easy to use and has a Command Line Interface (CLI) and configurable options to adjust the behavior of the tests depending on the user's necessity.
\end{itemize}

\subsection{OWASP ZAP}
OWASP ZAP (Zed Attack Proxy)\cite{ZAP} is one of the most important and popular tools in web \textit{pentesting}. This tool is free and open source, developed by Open Web Security Project (OWASP) and whose goal is to help security professionals or developers find vulnerabilities in web applications and test the security.

Features of OWASP ZAP:
\begin{itemize}
    \item Graphical and Command Line Interface: OWASP ZAP offers both a graphical user interface (GUI) and a command line interface (CLI), making it easy to use for the different types of users that may use it.
    \item Automatic Scanner: OWASP ZAP can automate a variety of security tests and generate reports with the findings.
    \item Spidering: OWASP ZAP can spider a web application to discover all its pages and resources, facilitating a more complete analysis.
\end{itemize}

\subsection{Burpsuite}
Burpsuite \cite{burpsuite} ranks among the most critical and widely used tools for web penetration testing. This tool was developed by PortSwigger and provides a wide range of functionalities for testing and detecting vulnerabilities.

Depending on the user interests, Burp Suite has several versions, that goes from a free one called Burp Suite Community Edition, with limited functionalities to perform basic security testing tasks to paid ones like Burp Suite Professional, with advanced features like automated scanning and technical support

\subsection{CCN and CCN-STIC 807}
\subsubsection{National Cryptology Center, CCN}
The National Cryptology Center (CCN), created in 2004, is an agency that belongs to the Spanish National Intelligence Center (CNI) and whose objective is to protect sensitive information, guarantee the security of information and communication technologies in Spain, and, apart from that, train specialist personnel in this field. This means that the CCN plays a fundamental role in Spain's cyber-defense and also in the protection of critical information systems of the government and other organizations \cite{ccn}.

The CCN conducts several activities, including the development and promotion of policies, guidelines, and security standards in the field of information and communication technologies. To this end, the CCN has developed the National Information and Communications Technology Security Scheme (CCN-STIC), which establishes a regulatory and technical framework for information security management in Spanish publishing administrations and some other public and private sector entities to improve cybersecurity.  The CCN-STIC offers a variety of resources, such as documents, manuals, and tools, which can be used as a reference to establish security measures in information systems and networks, as these resources address a wide variety of topics such as risk management, data protection, security in software development, and protection of critical systems among others\cite{ccn-stic}.

\subsubsection{CCN-STIC 807}
CCN-STIC 807\cite{ccn-stic-807} is a technical paper published by the National Cryptography Centre (CCN) in May 2022. This document is included in the regulations and guidelines to improve the security of information and communications systems and is focused on safety in the workplace. It provides recommendations and guidelines on how to configure, manage and secure workstations such as computers, laptops and other devices to ensure they are protected against potential security threats. 

The objective of this guide is to provide a description of the authorized cryptographic mechanisms for use in the National Security Scheme (ENS), detailing the necessary strength and the guarantees that they must comply with, according to the level of security required for each corresponding measure.

For doing so, the guide includes a list of authorized cryptographic mechanisms, their standards, the strength required, and some commonly used protocols according to the level of security needed. For example, RSA keys whose length is less than 2048 are stated as no recommended

\section{Architecture Overview}
\subsection{Global view}
The architecture has been carefully designed to ensure a clear and efficient identification of each task through the use of several modules. This modularity not only facilitates the understanding and maintenance of the code, but also allows the execution of tasks in parallel and with it, optimizing the performance and speed of the tool.  

In keeping with the fundamental objective of PTHelper, which is to minimise pentester interaction during the pentest phases, the system has been designed so that the only input required at the start of the evaluation is the initial information provided by the user. Thereafter, all other inputs and interactions that would normally require pentester intervention are handled automatically by PTHelper, reducing the numbers of steps the pentester must perform to obtain results and making the results more convenient to obtain.
In this work new functionalities for web \textit{pentesting} have been implemented in the PTHelper tool. That is why the architecture of PTHelper is divided in two main modules: host \textit{pentesting}, where the user provides an IP address and web \textit{pentesting}, where the user provides an URL.

In addition, the user can check the vulnerabilities found through the Command-line Interface (CLI) and when PTHelper finishes the execution, a report with the result is generated. 

\begin{figure}[h!]
    \centering
    \includegraphics[width=0.5\linewidth]{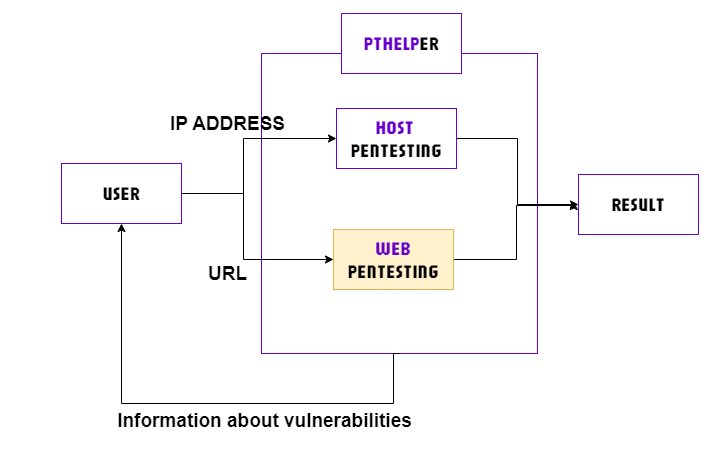}
    \caption{PTHelper: main architecture}
    \label{fig:main_architecture}
\end{figure}

Figure \ref{fig:main_architecture} gives an overview of the new PTHelper architecture, being the yellow block the new functionality added in this project.

\begin{figure}[h!]
    \centering
    \includegraphics[width=0.5\linewidth]{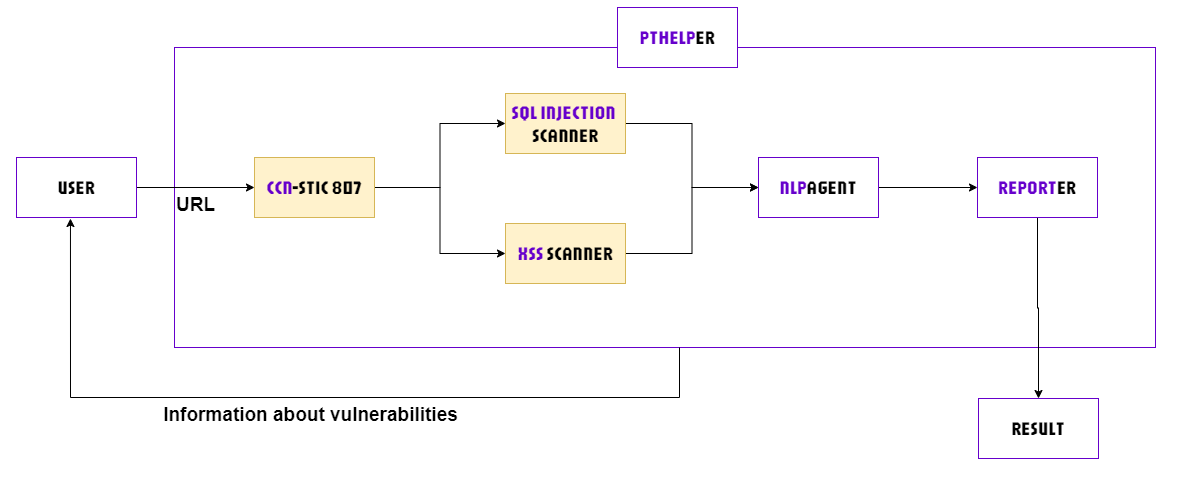}
    \caption{PTHelper: Web \textit{pentesting} architecture}
    \label{fig:web_architecture}
\end{figure}

Going into more detail in the web \textit{pentesting} module (Figure \ref{fig:web_architecture}), it can be divided into 5 different modules. The first module, called CCN STIC, is in charge of verifying the compliance of the web application with the CCN-STIC 807 guides. SQL Injection and XSS Scanner are the next two modules in charge of finding SQL injection and Cross-Site Scripting. This two modules are executed in parallel in order to improve the performance of the tool. The fourth module, NLPAgent is in charge of interfacing with an artificial intelligence agent to retrieve additional information about the vulnerabilities and how to prevent them. Finally, the fifth module takes care of compiling all the information from the previously executed modules and generating a report with it.

All the modules have been completely developed in this work except the NLPAgent and Reporter. Both modules already existed but they have been modified to incorporate the new web \textit{pentesting} functionality and new features explained in detail below.

In addition, the implemented system is parameterized, allowing to flexibly change both the commands and the tools used to discover vulnerabilities. This adaptive approach ensures that PTHelper can easily integrate and evolve with new technologies and security methods, keeping it always at the forefront of threat detection.

The following subsections provide a more detailed explanation of each module built into PTHelper for web \textit{pentesting}. In addition to that, it will be covered how these modules interact with each other and their collaboration to the full functionality of the tool. The last subsections will focus on the scope of the tool, the usage policy, and the management of the projects carried out using this tool.
\subsection{Initial checks}
Before modules are executed, an initial check is performed to ensure that the provided URL is valid and accessible, and that at least one of the arguments: \textit{ccn}, \textit{sql} or \textit{xss} is set to True. If this condition is not met, the part of the program related to web \textit{pentesting} stops immediately as it would be useless to continue the execution knowing that there will be no response and no scanning will be done. On the contrary, if the URL is accessible and at least one of the arguments is enabled, it means that the condition is satisfied and the program proceeds with its normal operation. 

Another check that is done is to verify that the URL passed as input uses HTTPS (and not HTTP). If it uses HTTP it means that it does not use SSL/TLS to encrypt the communication between the client and the server, so there are no security configurations to check. Therefore, if the URL uses HTTP, PTHelper does not execute the CCN-STIC 807 module, prints on the command line interface a message that to execute that module the provided URL shall use HTTPS and executes directly SQL injection and XSS modules.

\subsection{CCN-STIC 807}
This module is in charge of verifying the compliance of the web application with the CCN-STIC 807 standard. On ensuring compliance with established security standards, it helps in protecting the integrity, confidentiality and availability of information against possible threats and cyber-attacks. It is important to take into account that the CCN-STIC 807 may change over time as the technologies and threats in cybersecurity area are in constant evolution as the elements, among which are protocols, ciphers and elliptic curves, that nowadays are considered secure may become deprecated in a future due to the discovery of new vulnerabilities or to the emergence of new attack methods. Moreover, it is possible that new encryption methods and elements may be introduced which were not covered in previous versions of the CCN-STSIC 807 standard. That is why it is important to keep up to date with updates and versions of the standard to ensure that the most appropriate and up-to-date security measures are used and avoid implementing deprecated ones.  

\begin{figure}[h!]
    \centering
    \includegraphics[width=0.5\linewidth]{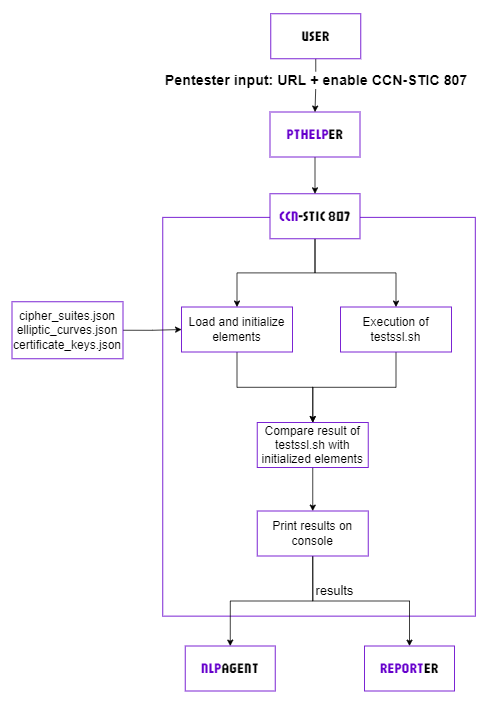}
    \caption{CCN-STIC 807 module operations and interactions with the other modules}
    \label{fig:ccn-stic_module}
\end{figure}
To facilitate the updating of the CCN-STIC 807 information in a fast and efficient way, all the security measures being checked are stored in a structured way in three different .json files so, if the CCN publishes a new release, only by changing the jsons files with the new security measures, it updates the whole project module (there is no need to change the source code of the program). The three .json files: \textit{cipher\_suites.json}, \textit{certificate\_key.json} and \textit{elliptic\_curves.json} store the elements included in the CCN-STIC 807 together with their characteristics. Examples of these files are included below.
\begin{verbatim}
 {
 "cipher_suite":
   "TLS_ECDHE_RSA_WITH_AES_256_GCM_SHA384",
 "tls": 1.2,
 "category":"R"
 }
\end{verbatim}
Example of element in cipher\_suites.json. In this case, the ciphersuite \textit{TLS\_ECDHE\_RSA\_WITH\_AES\_256\_GCM\_SHA384} with TLS version 1.2 is declared as recommended in CCN-STIC 807.
\begin{verbatim}
 {
 "key":"EC",
 "min_length": 256,
 "max_length": 283,
 "category":"R"
 }
\end{verbatim}
Example of element in certificate\_key.json. In this case, the certificate key \textit{EC} with length between 256 and 283 bits is declared as recommended in CCN-STIC 807.
\begin{verbatim}
 {
 "elliptic_curve":"brainpoolp256r1",
 "category":"R"
 }
\end{verbatim}
Example of element in elliptic\_curves.json. In this case, the elliptic curve \textit{brainpoolp256r1} is declared as recommended in CCN-STIC 807.

The field named category is present in all elements; this field can have 2 different values:
\begin{itemize}
    \item R (Recommended):  This category indicates that the mechanism is considered to be current and secure, and is recommended for use in new applications and systems as they offer an adequate level of long-term security. Technologies in this category are usually state-of-the-art and are considered suitable for protecting information in security environments.
    \item L (Legacy): This category refers to cryptographic technologies that have been widely used today, but offer an acceptable level of security only in the short term as they are insecure due to known vulnerabilities or advancing technologies. These technologies can only be used in scenarios where the threat is low to medium and the security level required by the system is also low to medium. In addition, their period of validity is established until December 31, 2025 unless another period is expressly indicated.
\end{itemize}

\begin{figure}[h!]
    \centering
    \includegraphics[width=0.5\linewidth]{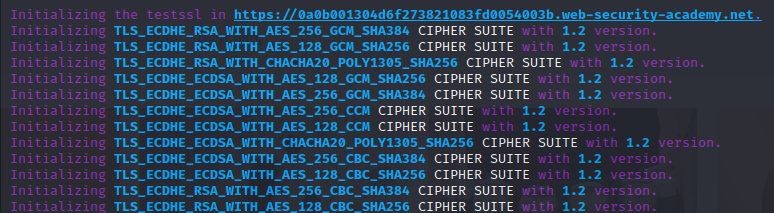}
    \caption{Initialization of the elements covered in CCN-STIC 807}
    \label{fig:console-init}
\end{figure}

This module starts (see Figure \ref{fig:ccn-stic_module}) by loading and initializing the security variables from the .json files (\textit{cipher\_suites.json, elliptic\_curves.json and certificate\_key.json}). While the data is being loaded, the user is able to see it as it is printed on the command line interface (see Figure \ref{fig:console-init}) thanks to the python class called Console \cite{console}

While the variables are initialized, testssl.sh is executed using a python module called \textit{subprocess}  \cite{subprocess} in order to check the page content. However, since testssl spends too much time, instead of executing the command just once in order to retrieve all the information, it is executed five times in parallel for different subsets of information. To do so, it passes as arguments the URL of the page and the specific argument for retrieving only the information of each element:
\begin{itemize}
    \item Protocols: Running testssl.sh with\textit{ -p} argument
    \item Ciphersuites: Running testssl.sh with\textit{ -E} argument
    \item Elliptic curves: Running testssl.sh with\textit{ -f} argument
    \item Certificate keys: Running testssl.sh with\textit{ -S} argument
    \item Vulnerabilities: Running testssl.sh with\textit{ -U} argument
\end{itemize}

Thanks to python parallelism and multiprocessing module\cite{multiprocessing}, the module is able to execute all the commands at the same time rather than sequentially executing them, thus improving performance. 

\begin{figure}[h!]
    \centering
    \includegraphics[width=0.5\linewidth]{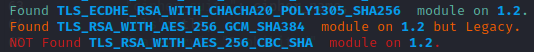}
    \caption{Example of results printed on the console}
    \label{fig:console-results}
\end{figure}

Once testssl.sh has finished running and the data has been loaded, it is compared with the previously initialized values from the .json files. For each element identified by testssl, it is printed on the Command Line Interface if it matches the standards established by CCN-STIC 807, indicating its status by means of colors (Figure \ref{fig:console-results}).
\begin{itemize}
    \item If an element is not in the standard, it will be highlighted in red.
    \item If an element has legacy category, it will be highlighted in orange
    \item If an element has recommended category, it will be highlighted in green
\end{itemize}

In addition to printing on the console, all the information collected will be stored in dictionaries so that the module can conclude by returning these data, which allow its later use to generate reports with the Reporter module and give information to the NLPAgent module. 

\subsection{SQL injection\label{sec:sql-injection}}
As mentioned above, one of the most common and dangerous attack vector is SQL injection and to mitigate it, a specialized module has been implemented to detect and report possible cases of SQL injection in web applications. Therefore, the objectives of this module are:
\begin{itemize}
    \item Detect vulnerabilities: Identify the places in the code where SQL queries can be maliciously manipulated in the  URL passed as input. Moreover, the module is able to check all the URLs that belong to the same domain of the input URL in order to find all the vulnerabilities in that domain.
    \item Report vulnerabilities: Provide detailed reports on the vulnerabilities found, including the location and nature of each vulnerability.
\end{itemize}

To make all this possible in the module and carry out this integration, the SQLMapAPI code has been modified in order to adjust and adapt it to the needs of PTHelper. In the context of the execution of external programs using the subprocess module, there is a significant limitation: only one call can be made at a time, as no persistent state is maintained between calls. To overcome this limitation, two modifications in the SQLMapAPI code have been essential to achieve this:
\begin{itemize}
    \item Modification of status and data management: Before the modification, it was necessary to execute two separate commands: first \textit{use \textless taskid\textgreater} and after that \textit{status} or \textit{data}, to retrieve the status of the task or the result of it. Thanks to the modification of the code, only one call is needed to retrieve the status or data of a specific task passing as argument the taskid. To sum up, this modification simplifies the process and reduces the number of interactions required to retrieve the status and data of a task.
    \item Deletion of not relevant information: The code has been purged of non-relevant information in this case, which has contributed to the improvement of system performance by reducing the amount of information transmitted, optimising the use of resources and speeding up the execution of tasks.
\end{itemize}

\begin{figure}[h!]
    \centering
    \includegraphics[width=0.4\linewidth]{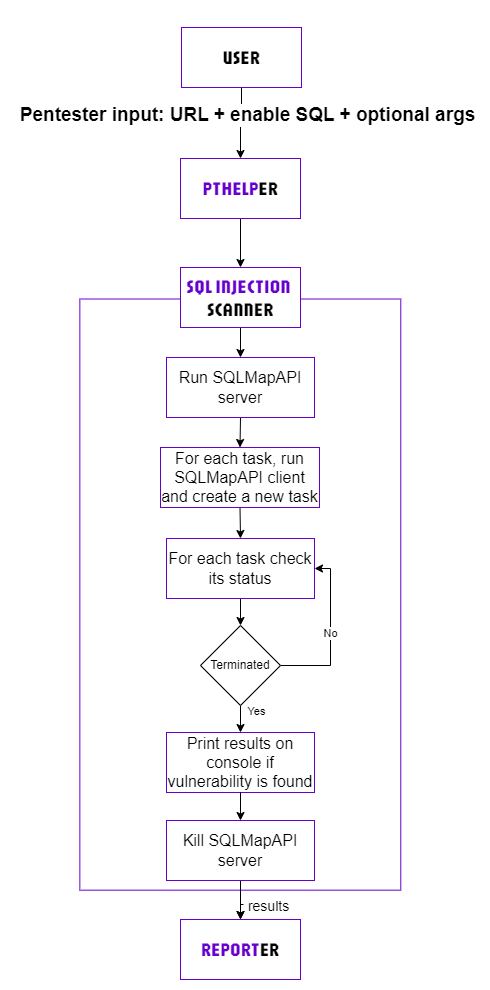}
    \caption{SQL Injection Scanner module operations and interactions with other modules}
    \label{fig:sql-scanner}
\end{figure}

The first fundamental step that the SQL Injection Scanner module performs  (see Figure \ref{fig:sql-scanner}) is the execution or initialization of the SQLMapAPI server to act as an intermediary between the future SQLMapAPI clients and the SQLMap engine that processes the tasks. This execution is also done with the subprocess module in order to start and control the server process in the background.

After the initialization of the SQLMapAPI server, the module uses the initial URL received to crawl other URLs belonging to the same domain, which will be analyzed for possible SQL injections. The HTML content of each URL is analyzed using BeautifulSoup\cite{beautifulsoup}, (python library for scraping information from web applications), extracting all the form elements of the page, including their attributes and fields, critical to performing SQL injection tests. After that, and in order to improve the analysis efficiency, ThreadPoolExecutor \cite{threadpool}, a Python subclass is used to execute several tasks in parallel. For each form element, a SQLMapAPI client is executed and a new task is launched to verify the presence of SQL Injection vulnerabilities. If there is not any form element in the URL, a new task is launched only with the URL, with the objective of detecting all the possible SQL injection vulnerabilities at other entry points such as parameters in the URL. This approach ensures complete coverage in the detection of possible SQL injection points, as both forms and URLs are analyzed directly.

PTHelper also allows to pass as argument several options to execute SQLMapAPI tasks with different parameters.
The risk and level parameters are used to adjust the aggressiveness and depth of the analysis to be performed and if not specified when running the program they take the SQLMap default values, which is 1 for both parameters.
In addition, if you wish to pass a specific username and password, the tool allows you to provide them as arguments and if at any time the tool finds a field corresponding to those values, it will replace the values of those fields with the values provided in the arguments. This functionality is very useful for verifying the security of applications that requires login.
Finally, in many forms there are hidden inputs, also called CSRF tokens. These tokens play a crucial role in the protection of web applications and although they are originally intended for Cross-Site Request Forgery attacks and not for SQL injection, they provide an additional layer of security by validating the authenticity of the requests and making SQL injection exploitation more difficult. When developing the tool, the presence of these tokens has been taken into account and therefore it is able to automatically extract them and include them in the requests sent to SQLMapAPI.

\begin{figure}[h!]
    \centering
    \includegraphics[width=0.5\linewidth]{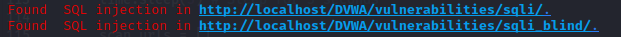}
    \caption{Found SQL injection message}
    \label{fig:sql-vuln-found}
\end{figure}

The next step is checking the status of each task and if the task has finished, check its response. If in the response there is any vulnerability, the relevant information of the vulnerability is stored and it prints a message stating that a SQL injection vulnerability has been found in that (Figure \ref{fig:sql-vuln-found}). Finally, when all the tasks have finished, the SQLMapAPI server is killed and the vulnerabilities found are returned.

It should be noted that SQLMapAPI has been used for this module but if in the future there is a better tool or in some case the pentester prefers to use another one, there would be no problem as it is parameterised and it would be enough to change the commands (and parameters if required) of the pthelper\_config.py file.
 
\subsection{XSS injection}
Another of the most common and dangerous attack vector is Cross-site Scripting and to mitigate it, a specialized module has been implemented to detect and report possible cases of XSS in web applications. Therefore, the objectives of this module are:
\begin{itemize}
    \item Detect vulnerabilities: Identify the places in the code where XSS queries can be maliciously manipulated in the  URL passed as input. Moreover, the module is able to check all the URLs that belong to the same domain of the input URL in order to find all the vulnerabilities in that domain.
    \item Report vulnerabilities: Provide detailed reports on the vulnerabilities found, including the location and nature of each vulnerability.
\end{itemize}
To make all this possible in the module and carry out this integration, XSpear, an automatic and advanced tool that detects and exploits cross-site scripting vulnerabilities has been used.

\begin{figure}[h!]
    \centering
    \includegraphics[width=0.4\linewidth]{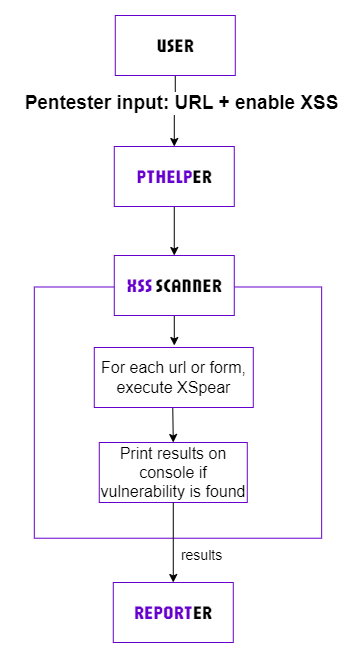}
    \caption{XSS Scanner module operations and interactions with other modules}
    \label{fig:xss-module}
\end{figure}

\begin{figure}[h!]
    \centering
    \includegraphics[width=0.5\linewidth]{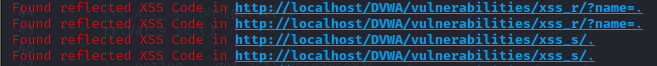}
    \caption{Found SQL injection message}
    \label{fig:xss-vuln-found}
\end{figure}

This module is similar to SQL Injection Scanner one but without the need of executing any server to act as an intermediary. The first step (Figure \ref{fig:xss-module}) of this module is to crawl all the URLs that belong to the same domain with the initial URL received. The HTML content of each URLs is also analyzed using BeautifulSoup, extracting all the form elements from the page. After extracting the form elements and in order to improve the analysis efficiency, ThreadPoolExecutor is used to execute several tasks in parallel, assigning a form element to each task. If there is not any form element in the URL, a new task is launched only with the URL, with the objective of detecting all the possible Cross-Site Scripting vulnerabilities at other entry points such as parameters in the URL. Moreover, if any vulnerability is detected, the relevant information of the vulnerability is stored and it prints a message stating that a XSS vulnerability has been found  (Figure \ref{fig:xss-vuln-found}) This approach ensures complete coverage in the detection of possible Cross-Site Scripting points, as both forms and URLs are analyzed directly.

It should be noted that XSpear has been used for this module but if in the future there is a better tool or in some cases the pentester prefers to use another one, there would be no problem as it is parameterised and it would be enough to change the commands of the \textit{pthelper\_config.py} file.

\subsection{NLPAgent}
In this part of the tool, the web \textit{pentesting} extension takes advantage of the previously implemented structure of the Natural Language Processing (NLP) module. This module employs artificial intelligence to perform various tasks, playing a crucial role in the aspects of the penetration testing process. The function performed by this module is to complete the evaluation report generated by the tool with information based on the evaluation context and several conversational prompts designed specifically for this purpose. The results generated range from executive reports to justifications of the severity of each identified vulnerability. The OpenAI API has continued to be used as an agent due to its ease of use and popularity.
\begin{figure}[h!]
    \centering
    \includegraphics[width=0.5\linewidth]{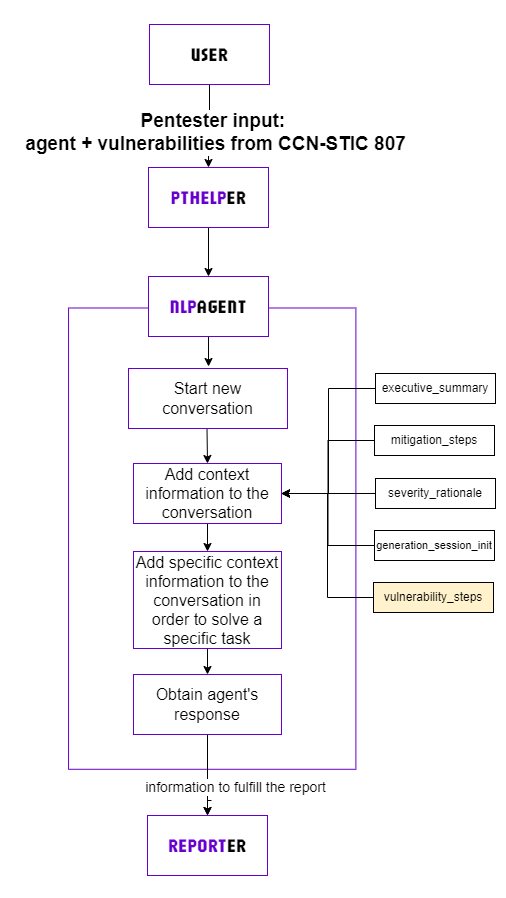}
    \caption{NLP agent module}
    \label{fig:nlp-agent}
\end{figure}

The agent adjusts its behavior and responses depending on the context given, letting the agent give detailed and specific responses for each section of the report. That is why, for the context of web penetration testing and following the previous methodology, a new function has been created to explain the vulnerabilities extracted from the CCN-STIC 807 module. In addition to explaining these vulnerabilities, it also provides concrete steps to mitigate them to avoid possible attacks. For doing so, apart from the prompts (executive\_summary, mitigation\_steps. severity\_rationale and generation\_session\_init) already established, a new one called vulnerability\_steps has been added (Figure \ref{fig:nlp-agent}).

\subsection{Reporter}
The Reporter module is a key component of this tool; it is the module in charge of consolidating and presenting the information obtained from the previously executed modules. The main function of this module is to transform the data and results received from each previous module and to unite everything in an understandable and useful format for the pentester, thus facilitating the interpretation and analysis of the findings.

\begin{figure}[h!]
    \centering
    \includegraphics[width=0.5\linewidth]{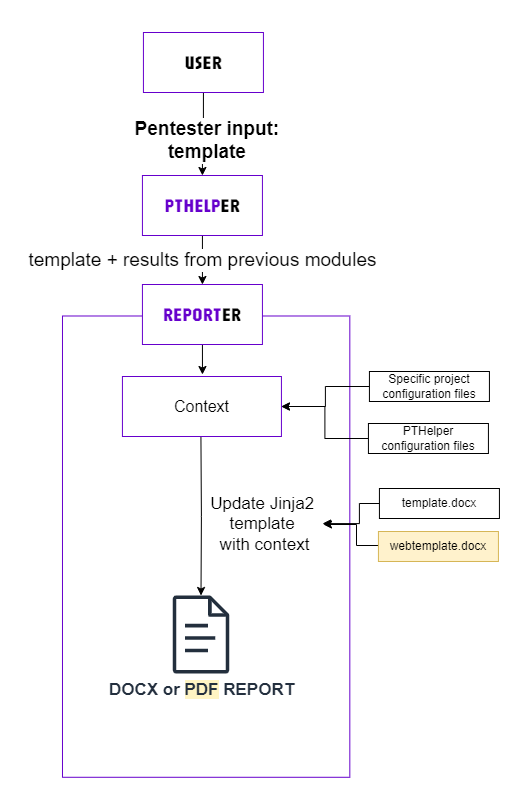}
    \caption{Reporter module}
    \label{fig:reporter}
\end{figure}

This module uses a new Jinja template \cite{jinja2} for web vulnerabilities (webtemplate.docx). The Jinja template allows to structure the report in a clear and organised way, defining the structure and format of the report and thus ensuring consistency and professionalism. The module (Figure \ref{fig:reporter}), by collecting the information from the previous modules, dumps them into the template, automatically filling in the corresponding sections with the data obtained and generating the report.
So, when executing the module, it will add to the web template the following information, providing a complete and wide vision of the security findings:
\begin{itemize}
    \item Summary of the encryption elements: It will be included a detailed summary of each encryption element and their classification according to CCN-STIC 807 standard, providing a security evaluation. Moreover, thanks to NLPAgent, the vulnerabilities found in the encryption elements will be described together with their mitigation steps.
    \item SQL injection cases:  All SQL injection cases encountered by the SQL Injection module will be listed along with their characteristics such as the URL where they are found, the payload used or a description.
    \item XSS cases: Similar to SQL injection ones, all XSS cases encountered by the XSS module will be listed along with their characteristics such as the URL where they are found, the payload used or a description.
\end{itemize}

In terms of formats, the report can be generated in 2 different formats: 
\begin{itemize}
    \item DOCX format: The report is created in Microsoft Word format (.docx) and was used in the first version of PTHelper to generate the report. This format is editable, allowing the pentester to customize the content as needed after running PTHelper.
    \item PDF format: The report is created in PDF format (using doc2pdf \cite{doc2pdf}), a new type of format added in this version  but works for both, normal pentesting and webpentesting. This format is widely preferred for final report distribution due to its resistance to modification and universal compatibility.
\end{itemize}

\subsection{Scope}
The new functionalities of the PTHelper tool have been developed to perform penetration testing (\textit{pentesting}) on web applications, focusing exclusively on identifying vulnerabilities within the domain specified by the user through the command line using \textit{--url }as an argument. Therefore, no tests will be performed on external URLs to ensure compliance with usage policies and applicable laws. 

\subsection{Usage Policy}
The use of the PTHelper tool must be in accordance with the ethical and responsible usage policies. This means that the user is responsible for ensuring that he/she has explicit permission to perform security testing on the target domain.

\section{Tool usage and information}
\subsection{Programming language}
As the original PThelper was fully developed in Python3\cite{python}, this programming language has been kept to add these new web \textit{pentesting} functionalities.

This programming language offers a wide variety of advantages, such as:
\begin{itemize}
    \item Simplicity and readability: Python3 has a clear and concise syntax, making the understanding of the code easier, improving collaboration and code maintenance
    \item Extensive Standard Library: Python3 has a large standard library that offers a great variety of tools and functionalities without the need to install external libraries.
    \item Portability: Python3 is compatible with a variety of platforms including Windows, macOs and Linux
    \item Active Community: As the Python community is active, it speeds up the development and troubleshooting process.
\end{itemize}

Apart of the inherent advantages there are:
\begin{itemize}
    \item Parallelism: Parallelism techniques can be implemented in Python3. This means that several tasks can be executed at the same time and improve the performance of the program by distributing the workload more efficiently and making the best use of the available resources.
    \item Easy integration with other tools: Python integrates easily with other tools and technologies used in \textit{pentesting} like APIs 
\end{itemize}
\subsection{Tool installation and dependencies}
The developed tool requires certain components and configurations to work properly. 
For the web \textit{pentesting} functionality, SQLMapAPI is used to run SQL injection tests, however, it is not necessary to install it as the code has been modified to better integrate it into the tool (as explained in Section \ref{sec:sql-injection} )and is included directly in the tool's code. For vulnerability detection, it is necessary to install XSpear using the command '\textit{gem install XSpear}'.

All the Python packages needed for the tool are listed in the \textit{requirements.txt} file and can be installed by running the following command from the terminal: ‘\textit{pip install -r requirements.txt}’. 
After installing all dependencies, the tool is installed using the following command from the root folder of the tool: '\textit{pip install -e .} '

Finally, the API key used for OpenAI and NVD API key will be added in pthelper\_config.py. In this file, you can also change the tools used to detect SQL injection or XSS according to the specific needs of the pentester (by default modified SQLMapAPI and XSpear are used). Following these steps, the tool is ready to use.
\subsection{Folder structure}
\begin{figure}[h!]
    \centering
    \includegraphics[width=0.5\linewidth]{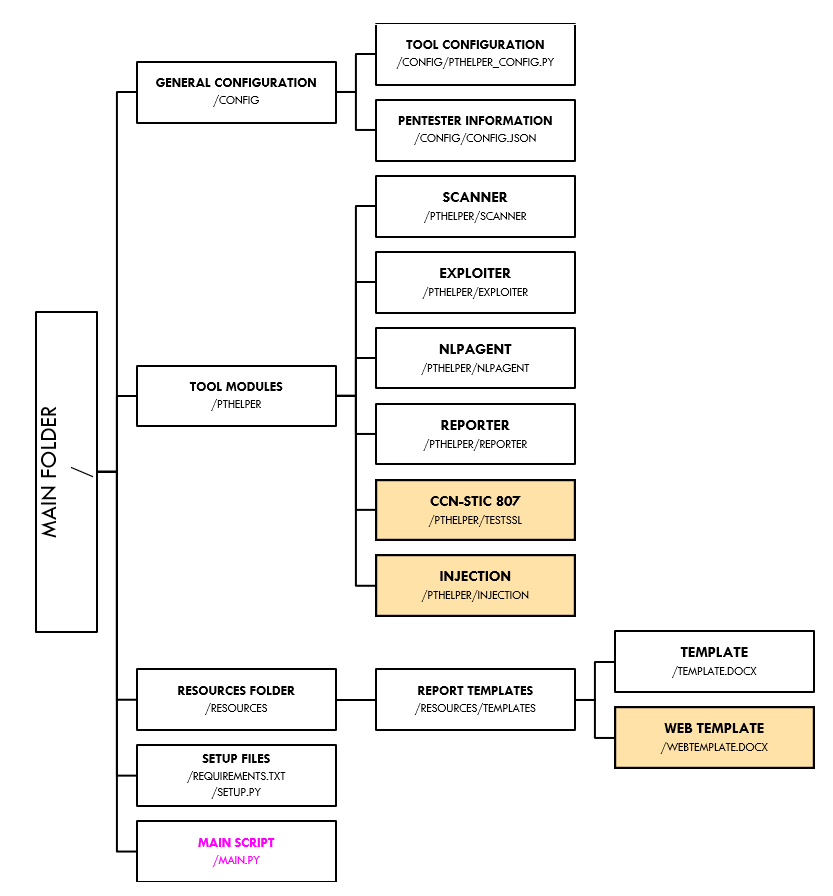}
    \caption{Structure of PTHelper}
    \label{fig:structure}
\end{figure}

The tool has been designed in a modular way to facilitate maintenance, scalability and customisation. The attached figure (Figure \ref{fig:structure}) provides a graphical view of this structure, making it easier to understand.
\begin{itemize}
    \item The config folder contains all configuration files. In pthelper\_config.py is where you find the global variables (like API keys) and the commands used for the injection modules (by default they use SQLMapAPI and XSpear). In addition to \textit{pthelper\_config.py}, another file called config.py is generated where the pentester information is stored, information that is then used to fill in the reports. 
    \item The PTHelper folder contains a folder for each module. In this case, two folders have been added, testssl folder for the ccn-stic 807 module and the injection folder for the SQL injection and XSS modules.
    \item In the resources folder apart from the previously added resources, a new resource has been added, the template for web \textit{pentesting}, called \textit{webtemplate.docx}
    \item In the root folder there are 2 different files, requirements.txt and setup.py for installing and configuring the tool properly.
    \item main.py file is the main entry point of the tool. Its function is to coordinate the execution of the modules, manage the data flow, and provide a user interface.
\end{itemize}
\subsection{Accepted parameters}
The tool is still usable via the Command Line Interface (CLI) and for the new functionality the necessary parameters would be:
\begin{itemize}
    \item --url: The URL to include in the assessment. The tool will analyze all the URLs from that website in order to guarantee complete coverage (but without including other external sites)
    \item --ccn: Boolean parameter specifying whether the CCN-STIC 807 compliance check is performed. By default is set to True.
    \item --sql: Boolean parameter specifying whether detection of possible SQL injections will be performed. By default is set to True
    \item --level: Integer parameter specifying the intensity of testing (1-5, where 5 is the most exhaustive) By default is set to 1
    \item --risk:  Integer parameter specifying the type of testing based on risk (1-3, where 3 is the most risky). By default is set to 1
    \item --xss:  Boolean parameter specifying whether detection of possible Cross-site scripting (XSS) will be performed. By default is set to True
    \item --user: String parameter specifying the user value for SQL injection. Optional parameter
    \item --password: String parameter specifying the password value for SQL injection. Optional parameter
    \item --agent: Type of NLP agent used.  At the moment only chatgpt agent is available
    \item --reporter: Type of reporter used to generate the report.  At the moment only docxtpl (Docx Jinja3 framework)  and pdf are available
    \item --project: Let the user specify the name of the project where the results will be stored. If the project does not exist, the tool will ask for some information like the target organization and will automatically create a project and a folder to store the results
\end{itemize}
\section{Testing PTHelper new functionalities}
To verify and demonstrate the functionality of the tool, extensive testing has been carried out using vulnerable web applications. These tests have been essential to assess the effectiveness and accuracy of the new modules created. 
\subsection{DVWA}
DVWA (Damn Vulnerable Web Application) is a PHP/MySQL web application designed to be insecure and vulnerable. It has been created with the aim of testing the skills and tools of pentests in a legal environment. This application allows pentesters to practice attack techniques and vulnerability exploitation in a controlled environment \cite{dvwa}.

This application includes both SQL and XSS injection, being a good option to test our tool, PTHelper.

\subsubsection{Expected behavior when running the tool with DVWA on HTTP protocol}
When running the tool with the web application running on http://localhost, it is expected that the first thing to be checked is the accessibility of the website. Although it is accessible, it is running on the HTTP protocol and not on HTTPS so the CCN-STIC 807 module will not be executed and will proceed directly to execute the SQL injection and XSS modules.
As for the SQL injection module, PTHelper is expected to be able to detect normal SQL injections as well as blind SQL injections. In addition, the XSS module is expected to identify reflected and stored XSS vulnerabilities. Finally all the information gathered will be dumped into the report file
\subsubsection{Results obtained}
The results obtained are as expected with a certain difference, PTHelper is able to detect SQL injection and XSS cases, in fact, it detects one case more than expected in SQL injection. However, by testing manually you can see that this extra case is a false positive.

Although PTHelper displays a false positive, this is not the fault of the tool itself. Rather, since it uses SQLMap underneath, the false positive is originating from SQLMap. 
\subsection{PortSwigger}
PortSwigger, the company behind BurpSuite, has provided hands-on learning environments called labs. The purpose of these labs is to help pentesters improve their skills in penetration testing and web application security in a lawful and ethical manner.

These labs, running on the HTTPS protocol, have a variety of vulnerable web application scenarios in controlled and secure environments to simulate real-world situations where SQL injection, XSS and other vulnerabilities can be found.
\subsubsection{Expected behavior}
When running the tool passing as argument the  of one of the Portswigger labs, the tool will check the accessibility of the  and its protocol. As it runs on HTTPS protocol and CCN-STIC 807 is enabled, the tool will check the encryption elements and classify them using  CCN-STIC 807 as a reference. 
After finishing the CCN-STIC 807 module, both SQL injection and XSS modules start and detect the vulnerabilities that appear on the page.
Finally, the report is generated and fulfilled with the vulnerabilities founded
\subsubsection{Results obtained}
The results obtained are the expected ones.

\begin{figure}[h!]
    \centering
    \includegraphics[width=0.5\linewidth]{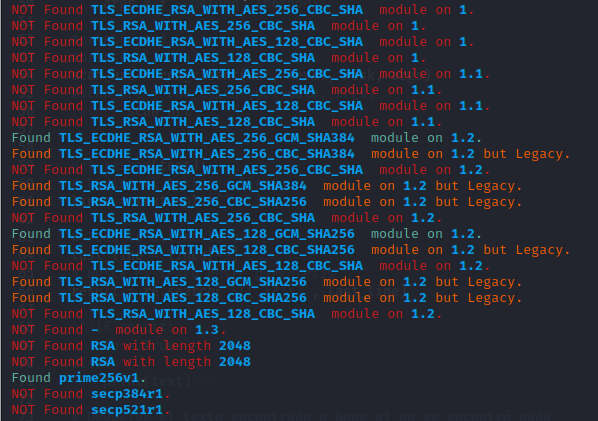}
    \caption{Result of CCN-STIC 807 module in the lab}
    \label{fig:result-ccnstic}
\end{figure}

For example, for the lab \textit{\textbf{SQL injection vulnerability in WHERE clause allowing retrieval of hidden data}}\cite{sql-lab} PTHelper executes the CCN-STIC 807 module to classify the encryption elements used (see Figure \ref{fig:result-ccnstic}).

\begin{figure}[h!]
    \centering
    \includegraphics[width=0.5\linewidth]{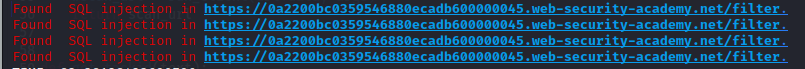}
    \caption{Result of SQL injection module in the lab}
    \label{fig:result-sql}
\end{figure}

After that, the module for scanning SQL injection discover one in  https://0a2200bc0359546880ecadb600000045.web-security-academy.net/filter  with query 'category=Accessories' (see Figure \ref{fig:result-sql}). Note that it appears that the injection 4 times as 4 different types of SQL injection can be used).

\begin{figure}[h!]
    \centering
    \includegraphics[width=0.5\linewidth]{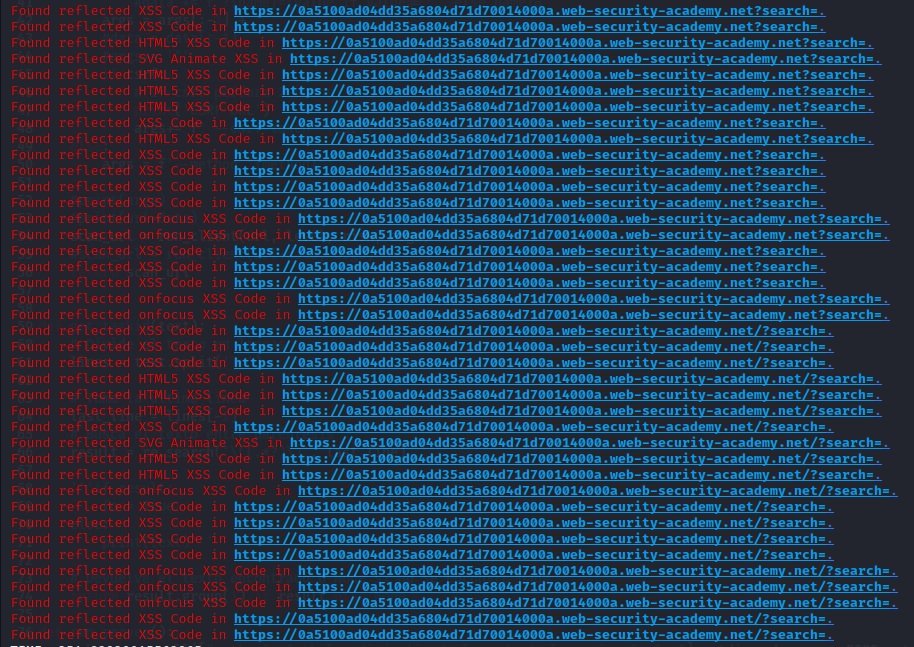}
    \caption{Result of XSS module in the lab}
    \label{fig:result-xss}
\end{figure}

Another example could be \textbf{\textit{Reflected XSS into HTML context with nothing encoded}}\cite{xss-lab}, PTHelper executes the CCN-STIC 807 and retrieves the same classification as the previous one (see Figure \ref{fig:result-ccnstic}). After that, the XSS module discovers a Reflected XSS in https://0a5100ad04dd35a6804d71d70014000a.web-security-academy.net?search= with parameter search.  (see Figure \ref{fig:result-xss}. Note that it appears the xss several times as different types of payloads can be used).

\subsection{Juice Shop}
Juice Shop is a deliberately insecure web application designed to practice penetration testing. Developed by OWASP, Juice Shop simulates a juice shop with a wide range of vulnerabilities including both PTHelper tested, SQL injection and XSS  \cite{juice-shop}.

However, it has a difference with the other web applications, it is developed using JavaScript and Angular. As traditional HTML parsing tools, such as BeautifulSoup, are not able to handle dynamically generated JavaScript content, the tool is not able to find forms and s that belong to the same domain as the initial one. Therefore, it is not able to find SQL injection nor XSS web vulnerabilities.

The code is available at \url{https://github.com/MariaOlivaress/PTHelper} and some videos and reports of the tests are available at \url{https://drive.google.com/drive/folders/1ezlC5vwfYRqw4mpjk95t-yrFn5JDGquk?usp=drive_link}
\section{Conclusion}
The addition of web \textit{pentesting} capabilities to the PTHelper tool means an advance and improvement in its functionality and usefulness as with this integration the tool offers pentesters a solution for the evaluation and improvement of web application security. 
Benefits
\begin{itemize}
    \item The tool is able to automatically detect SQL and XSS injection web vulnerabilities, which are very common in web applications. In addition, the pentester can customise vulnerability searches with parameters such as risk or level.
    \item The tool is able to identify the encryption elements used and classify them using CCN-STIC 807 as a reference.
    \item The tool has been developed in a modular way, making it easy to modify or add new functionalities. In addition, it is possible to customize the configuration, adapting the tool to the needs of the pentester and being able to use the best tool available for each type of test.
    \item The tool has been developed using parallelism to improve performance. SQL injection and XSS modules are executed at the same time.
    \item The tool generates reports in docx and pdf formats, presenting the results in a clear way in order to facilitate the interpretation of the findings and the implementation of corrective measures.
    \item Finally, the tool is easy to install, accessible and easy to use for pentesters. It is only necessary to run the tool, the pentester does not need to perform any further actions, the tool does it by itself.
\end{itemize}
In order to further improve and expand the capabilities of the tool, several lines of future development are proposed:
\begin{itemize}
    \item Detection of more types of web vulnerabilities: Incorporating the detection of a wider range of vulnerabilities such as CSRF, OS Command Injection or any that appears in OWASP Top 10.
    \item Analysis of JavaScript pages: Develop or improve vulnerability detection modules in applications that rely primarily on JavaScript and/or frameworks such as React or Angular. Implement  both static and dynamic analysis of JavaScript code to identify more vulnerabilities.
    \item Mobile Application Vulnerability Detection: Extend the functionality of the tool to include vulnerability detection in mobile applications (iOS and Android).
\end{itemize}

\bibliographystyle{IEEEtran}

\bibliography{pthelperweb}

\end{document}